\documentclass[twocolumn, superscriptaddress, preprintnumbers, amsmath, amssymb]{revtex4}

\allowdisplaybreaks
\allowdisplaybreaks[4]

\usepackage{CJK}
\usepackage{graphicx}
\usepackage{dcolumn}
\usepackage{bm}
\usepackage[dvipdfm,
            pdfstartview=FitH,
            CJKbookmarks=true,
            bookmarksnumbered=true,
            bookmarksopen=true,
            colorlinks=true,
            pdfborder=001,
            linkcolor=blue,
            anchorcolor=blue,
            citecolor=blue
            ]{hyperref}

\begin{document}


\title{$g$-factor and static quadrupole moment for the
wobbling mode in $^{133}$La}

\author{Q. B. Chen}\email{qbchen@pku.edu.cn}
\affiliation{Physik-Department, Technische Universit\"{a}t
M\"{u}nchen, D-85747 Garching, Germany}

\author{S. Frauendorf}\email{sfrauend@nd.edu}
\affiliation{Physics Department, University of Notre Dame, Notre
Dame, IN 46556, USA}

\author{N. Kaiser}\email{nkaiser@ph.tum.de}
\affiliation{Physik-Department,
             Technische Universit\"{a}t M\"{u}nchen,
             D-85747 Garching, Germany}

\author{Ulf-G. Mei{\ss}ner}\email{meissner@hiskp.uni-bonn.de}
\affiliation{Helmholtz-Institut f\"{u}r Strahlen- und Kernphysik and
             Bethe Center for Theoretical Physics, Universit\"{a}t Bonn,
             D-53115 Bonn, Germany}
\affiliation{Institute for Advanced Simulation,
             Institut f\"{u}r Kernphysik and J\"{u}lich Center for Hadron Physics,
             Forschungszentrum J\"{u}lich,
             D-52425 J\"{u}lich, Germany}
\affiliation{Ivane Javakhishvili Tbilisi State University, 0186
             Tbilisi, Georgia}

\author{J. Meng}\email{mengj@pku.edu.cn}
\affiliation{State Key Laboratory of Nuclear Physics and Technology,
             School of Physics, Peking University,
             Beijing 100871, China}
\affiliation{Yukawa Institute for Theoretical Physics,
             Kyoto University,
             Kyoto 606-8502, Japan}
\date{\today}

\begin{abstract}

The $g$-factor and static quadrupole moment for the wobbling mode in
the nuclide $^{133}$La are investigated as functions of the spin $I$
by employing the particle rotor model. The model can reproduce the
available experimental data of $g$-factor and static quadrupole
moment. The properties of the $g$-factor and static quadrupole
moment as functions of $I$ are interpreted by analyzing the angular
momentum geometry of the collective rotor, proton-particle, and
total nuclear system. It is demonstrated that the experimental value
of the $g$-factor at the bandhead of the yrast band leads to the
conclusion that the rotor angular momentum is $R\simeq 2$.
Furthermore, the variation of the $g$-factor with the spin $I$
yields the information that the angular momenta of the
proton-particle and total nuclear system are oriented parallel to
each other. The negative values of the static quadrupole moment over
the entire spin region are caused by an alignment of the total
angular momentum mainly along the short axis. Static quadrupole
moment differences between the wobbling and yrast band originate
from a wobbling excitation with respect to the short axis.

\end{abstract}

\maketitle


The collective motions of a triaxially deformed nucleus, that is shaped like
an ellipsoid with three principal axes of inertia, have attracted a lot of
attention in nuclear structure physics over the last years. When
such a nucleus rotates, the lowest energy state for a given angular momentum
$I$ (called yrast state) corresponds to a uniform rotation about the principal
axis with the largest moment of inertia (MoI). At a slightly higher excitation
energy, this axis can execute a precession motion (in the form of harmonic
oscillation) about the space-fixed angular momentum vector. This describes the
phenomenon of the so-called wobbling motion that has been first proposed by
Bohr and Mottelson in the 1970s~\cite{Bohr1975}. Since this collective mode is
a rotation about a principal axis, the related energy spectra come as a series
of rotational $\Delta I=2$ bands in which the signature of the bands alternates
with increasing number of oscillation quanta $n$. The electric quadrupole
transitions with $\Delta I=1$ and $n\to n-1$ are induced by a
wobbling motion of the entire charged rigid body, and thus get
collectively enhanced.

Recent studies of the nuclear wobbling motion have been triggered by the
novel concepts of \textit{transverse} wobbling (TW) and \textit{longitudinal}
wobbling (LW) proposed by Frauendorf and D\"{o}nau~\cite{Frauendorf2014PRC}.
These authors have classified the wobbling modes in the presence of a high-$j$
quasi-particle according to the relative orientation of the angular momentum
of the quasi-particle $\bm{j}_p$ and the principal axis with the largest MoI,
which is usually the medium axis. If this relative orientation is perpendicular,
one speaks of a \emph{transverse} wobbling mode, and the corresponding wobbling
energy decreases with the spin $I$. It has been observed experimentally
for the nuclei $^{161}$Lu~\cite{Bringel2005EPJA}, $^{163}$Lu~\cite{
Odegaard2001PRL, Jensen2002PRL}, $^{165}$Lu~\cite{Schonwasser2003PLB},
$^{167}$Lu~\cite{Amro2003PLB}, and $^{167}$Ta~\cite{Hartley2009PRC} in the
$A\approx 160$ mass region, for the nuclides $^{135}$Pr~\cite{Matta2015PRL,
Sensharma2019PLB} and $^{130}$Ba~\cite{Petrache2019PLB, Q.B.Chen2019PRC_v1} in
the $A\approx 130$ mass region, and for the odd-neutron nuclide
$^{105}$Pd~\cite{Timar2019PRL} in the $A\approx 100$ mass region. If the
relative orientation is parallel, one speaks of a \emph{longitudinal} wobbling
mode, where the wobbling energy increases with the spin $I$. The experimental
evidence for longitudinal wobbling is, however, rare and has only been reported
very recently for the nuclides $^{133}$La~\cite{Biswas2019EPJA} and
$^{187}$Au~\cite{Sensharma2020PRL}.

The increase of the wobbling energy with the spin $I$ for the nucleus $^{133}$La
is quite unexpected, because the wobbling mode is based on the configuration
$\pi(1h_{11/2})^1$ with an orientation of the $h_{11/2}$ proton along the
short axis~\cite{Biswas2019EPJA}. The same $h_{11/2}$ proton configuration
applies to the isotones $^{135}$Pr~\cite{Matta2015PRL, Sensharma2019PLB}
and $^{131}$Cs~\footnote{http://www.nndc.bnl.gov/ensdf/.}, which both
show a decrease of the wobbling energy with the spin $I$, and this behavior
is actually a hallmark of the transverse wobbling mode. The authors of
Ref.~\cite{Biswas2019EPJA} have explained the unexpected increase of the
wobbling energy with spin $I$ for $^{133}$La as follows. Like the isotones,
this nucleus is triaxially deformed and thus it features the wobbling mode.
The triaxial deformation parameters $\beta=0.16$ and $\gamma=26^\circ$
are supported by tilted axis cranking calculations and the increase of the
wobbling energy with spin $I$ is attributed to nearly equal MoIs with respect
to the short ($s$-) axis and medium ($m$-) axis ($\mathcal{J}_s\simeq
\mathcal{J}_m$). The $m$-axis is no longer preferred for alignment with the
collective rotor angular momentum $\bm{R}$, which has now a larger component
along the $s$-axis. The mechanism underlying the enlarged $\mathcal{J}_s$-value
is attributed to the gradual alignment of a pair of positive-parity $(gd)$
protons with the $s$-axis. In the calculations of Ref.~\cite{Biswas2019EPJA},
this mechanism is taken into account by introducing a spin-dependent MoI for
$\mathcal{J}_s$. Such a scenario is obviously more complex than
the classification scheme suggested in Ref.~\cite{Frauendorf2014PRC}, which
assumes $\mathcal{J}_s<\mathcal{J}_m$. Therefore, the situation in $^{133}$La
corresponds to an intermediate coupling scheme.

Very recently, in Ref.~\cite{Laskar2020PRC} the $g$-factor and the static
(spectroscopic) quadrupole moment (SQM) were measured for the bandhead state
(an $11/2^-$ isomeric state) of the yrast band of $^{133}$La. The obtained
$g$-factor is $g = 1.16 \pm 0.07$ and SQM is $|Q| = 1.71 \pm 0.34~e\textrm{b}$.
On the theoretical side, Monte Carlo shell-model (MCSM) calculations
gave $g=1.16$ and provided the information that the dominant configuration
of the $11/2^-$ isomeric state is $\pi(1h_{11/2})^1$. At the same time,
these calculations predicted $Q = -1.25~e\textrm{b}$. The distribution of the
quadrupole moment expectation values obtained with shell-model wave functions
for the $11/2^-$ state indicates a triaxial shape with deformation
parameters $\beta\sim 0.16$ and $\gamma\sim 20^\circ$~\cite{Laskar2020PRC},
which is consistent with the wobbling interpretation. These new results motivate
us to investigate the $g$-factor and SQM as functions of the spin $I$
in the wobbling motion by taking the case of $^{133}$La as the first
example.

Our calculations are carried out with the particle rotor model (PRM), which
has been used widely for describing wobbling bands and has achieved much
success in this respect~\cite{Odegaard2001PRL, Jensen2002PRL, Hamamoto2002PRC,
Hamamoto2003PRC, Frauendorf2014PRC, W.X.Shi2015CPC, Streck2018PRC, Timar2019PRL,
Sensharma2019PLB, Biswas2019EPJA, Q.B.Chen2019PRC_v1, Sensharma2020PRL,
Q.B.Chen2020arXiv}. In Ref.~\cite{Biswas2019EPJA}, the PRM (there called
``quasi-particle plus triaxial rotor model'') could reproduce well the experimental
energy spectra and wobbling energies together with the electromagnetic transition
probability ratios $B(M1)_{\textrm{out}}/B(E2)_{\textrm{in}}$ and
$B(E2)_{\textrm{out}}/B(E2)_{\textrm{in}}$ for the wobbling bands
in $^{133}$La. In this work we use the same triaxial deformation
parameters $\beta=0.16$ and $\gamma=266^\circ$ as in Ref.~\cite{Biswas2019EPJA}.
With this chosen value of $\gamma$, the 1-axis, 2-axis, and 3-axis
are the conventional $s$-axis, $l$-axis, and $m$-axis of the triaxially deformed
ellipsoid, respectively. The MoIs of the nuclear core are taken as
$\mathcal{J}_m=15.33~\hbar^2/\textrm{MeV}$, $\mathcal{J}_l=2.92~\hbar^2/\textrm
{MeV}$, and $\mathcal{J}_s=[9.125 +0.657(I-j)]~\hbar^2/\textrm{MeV}$ for the $m$-,
$l$-, and $s$-axes, respectively~\cite{Biswas2019EPJA}. Here, $I$ and $j=11/2$
are the quantum numbers related to the total angular momentum $\bm{I}$ and
the proton angular momentum $\bm{j}_p$.

In the following, the methods to calculate the $g$-factor and SQM are given.
For an odd-mass nuclear system the rotor angular momentum $\bm{R}$ and the
(proton) particle angular momentum $\bm{j}_p$ are coupled to the total spin
$\bm{I}$ as
\begin{align}
 \bm{R}+\bm{j}_p=\bm{I}~.
\end{align}
The magnetic moment $\mu$ of this system is calculated from the
(rotational) wave function $|I,M=I\rangle$, with $M$ the quantum
number related to the projection of $\bm{I}$ onto the $z$-axis of
the laboratory frame, as follows
\begin{align}
 \label{eq4}
 \mu=gI=\langle II|g\hat{I}_z|II\rangle=\langle II|g_p\hat{j}_{pz}+g_R \hat{R}_{z}|II\rangle~,
\end{align}
where $\hat{I}_{z}$, $\hat{j}_{pz}$, and $\hat{R}_z$ are the
$z$-components of the respective angular momentum operators.
Moreover, $g_p$ and $g_R$ are the gyromagnetic ratios of the
proton-particle and the core, while the output quantity $g$ refers
to the total nuclear system. In the present study, we use the values
$g_R=Z/A=0.43$ for the rotor and $g_p= 1.21$ for the $h_{11/2}$
valence-proton, in which the spin $g$-factor $g_s=3.35$ has been
reduced to 0.6 times that of a free proton~\cite{Bohr1975}. The
possible modification of $g_R$ by the ($dg$) quasi-proton alignment
can be neglected, because the spin contribution to the $g$-factor is
small for normal-parity single-particle states.

By using the generalized Land\'{e} formula, the matrix element in
Eq.~(\ref{eq4}) can be expressed through scalar products of angular momentum
operators as
\begin{align}
 \mu=\frac{\langle II|g_p \bm{j}_p\cdot\bm{I} + g_R \bm{R}\cdot\bm{I}|II\rangle}{I(I+1)}
 \langle II|\hat{I}_z|II\rangle~.
\end{align}
After some rearrangement of terms, the $g$-factor of the total nuclear system
is given by
\begin{align}
g&=\frac{\langle g_p \bm{j}_p\cdot\bm{I} + g_R \bm{R}\cdot\bm{I} \rangle}{I(I+1)}\\
\label{eq3}
 &=g_R+(g_p-g_R)\frac{\langle \bm{j}_p\cdot\bm{I} \rangle}{I(I+1)}\\
\label{eq1}
 &=g_R+(g_p-g_R)\frac{j(j+1)}{I(I+1)}+(g_p-g_R)\frac{\langle \bm{j}_p\cdot\bm{R} \rangle}{I(I+1)}\\
\label{eq2}
 &=\frac{1}{2}\Big[(g_p+g_R)+(g_p-g_R)\frac{j(j+1)-\langle \bm{R}^2\rangle}{I(I+1)}\Big]~.
\end{align}
From Eqs.~(\ref{eq3}) and (\ref{eq1}), one sees that the $g$-factor reflects
the relative orientations between $\bm{j}_p$, $\bm{R}$, and $\bm{I}$.
In particular, one gets $g=g_R$ in the case  $\bm{j}_p\perp \bm{I}$, while
\begin{align}
 g=g_R+(g_p-g_R)\sqrt{\frac{j(j+1)}{I(I+1)}}
\end{align}
in the case $\bm{j}_p \parallel \bm{I}$, and
\begin{align}
 g=g_R+(g_p-g_R)\frac{j(j+1)}{I(I+1)}
\end{align}
in the case  $\bm{j}_p\perp \bm{R}$. It is worth mentioning here that
the $g$-factor has been used to investigate the angular momentum coupling
scheme in chiral doublet bands~\cite{Frauendorf1997NPA, Grodner2018PRL}. According to
Eq.~(\ref{eq2}), one can get from the resulting $g$-factor also information
about the rotor angular momentum, via the expectation value $\langle \bm{R}^2\rangle$.

The SQM gives a measure of the nuclear charge distribution associated
with the collective rotational motion, and it is calculated
as~\cite{Bohr1975, Ring1980book}
\begin{align}
 Q(I)=\langle II|\hat{Q}_{20}|II\rangle~,
\end{align}
where the quadrupole moment operator in the laboratory frame $\hat{Q}_{20}$
is obtained from the intrinsic quadrupole moments $Q_{2\nu}^\prime$ by
multiplication with Wigner D-functions:
\begin{align}\label{eq7}
 \hat{Q}_{20}=\sum_{\nu} D_{0,\nu}^{2} Q_{2\nu}^\prime~.
\end{align}
The 5 intrinsic quadrupole moments are $Q_{20}^\prime =Q_0^\prime\cos\gamma$,
$Q_{21}^\prime =Q_{2-1}^\prime=0$, $Q_{22}^\prime =Q_{2-2}^\prime=Q_0^\prime\sin\gamma
/\sqrt{2}$,  where $Q_0^\prime$ is an empirical quadrupole moment that is related
to the axial deformation $\beta$ by  $Q_0^\prime=3R_0^2Z\beta/\sqrt{5\pi}$,
with $Z$ the proton number and $R_0=1.2\,{\rm fm}\,A^{1/3}$.

In Ref.~\cite{Q.B.Chen2020arXiv_v1}, it is shown that the SQM can be also
calculated from expectation values of the squared total angular momentum
components along the three principal axes $\langle \hat{I}_k^2\rangle$ as
\begin{align}
 \label{eq5}
 Q(I) &=Q_0(I)+Q_2(I)~,\\
 \label{eq10}
 Q_0(I)&=\frac{3\langle \hat{I}_3^2\rangle-I(I+1)}{(I+1)(2I+3)}Q_0^\prime \cos\gamma~,\\
 \label{eq11}
 Q_2(I)&=\frac{\sqrt{3}(\langle \hat{I}_1^2\rangle-\langle \hat{I}_2^2\rangle)}
        {(I+1)(2I+3)}Q_0^\prime\sin\gamma~.
\end{align}
Therefore, the SQM can provide information about the orientation of
the total nuclear system relative to the principal axes frame.
Note that in the prolate case $\gamma=0^\circ$, when $\langle \hat{I}_3\rangle=K$
is a good quantum number, the part $Q_2(I)$ vanishes and $Q(I)$ becomes
\begin{align}
  Q(I)&=\frac{3K^2-I(I+1)}{(I+1)(2I+3)}Q_0^\prime~.
\end{align}
Using this formula, a deformation parameter of $\beta=0.28\pm 0.10$ has been extracted
in Ref.~\cite{Laskar2020PRC} under the assumption $K=1/2$ from the measured SQM
$|Q| = 1.71 \pm 0.34~e\textrm{b}$ for the $11/2^-$ isomeric state at the bandhead of
$^{133}$La.

\begin{figure}[!ht]
  \begin{center}
    \includegraphics[width=7.0 cm]{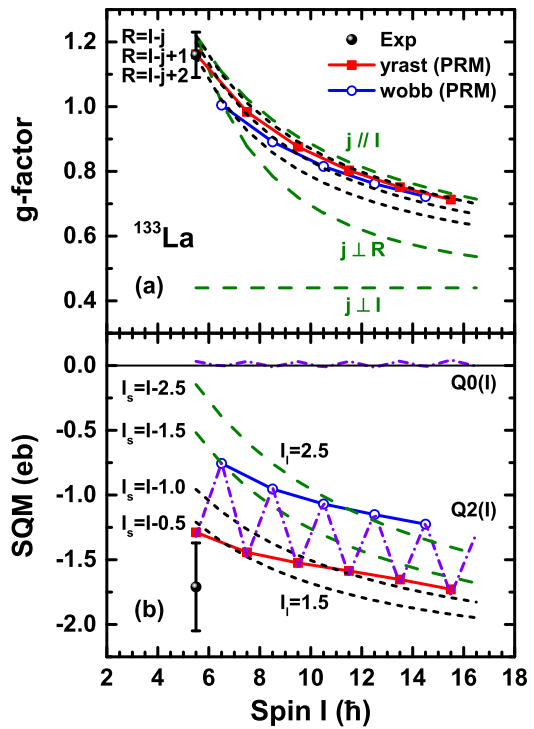}
    \caption{$g$-factors (a) and static quadrupole moments (b) as functions of the
    spin $I$ calculated in the PRM for the yrast and wobbling bands
    in $^{133}$La in comparison to the available data~\cite{Laskar2020PRC}. (a)
    The dashed lines correspond to  parallel and perpendicular couplings of the
    angular momenta of the proton and rotor. The short dashed lines represent the
    formula in Eq.~(\ref{eq2}) evaluated with different rotor angular momentum
    quantum numbers: $R=I-j$, $I-j+1$, and $I-j+2$. (b) The dashed-dot lines for
    $Q_0(I)$ and $Q_2(I)$ follow from Eqs.~(\ref{eq10}) and (\ref{eq11}). The dashed
    and short-dashed lines show $Q_2(I)$ calculated by Eq.~(\ref{eq11}) using the
    values of $I_s$ and $I_l$ as specified in the figure.}\label{fig1}
  \end{center}
\end{figure}

In Fig.~\ref{fig1}, we show the $g$-factors and SQMs as functions of
the spin $I$ as calculated in the PRM for states in the yrast and
wobbling bands of $^{133}$La in comparison with the available experimental
data~\cite{Laskar2020PRC}.

The PRM reproduces well the experimental $g$-factor at the bandhead of the yrast band.
The theoretical prediction $g= 1.16$ is in excellent agreement with the
experimental value $g=1.16 \pm 0.07$. The calculated $g$-factors decrease
with increasing spin $I$. This feature comes mainly from the denominator $I(I+1)$
in Eqs.~(\ref{eq3})-(\ref{eq2}). Over the entire spin region the $g$-factors for
states in the yrast band are larger than those for states in the wobbling band.
According to Eq.~(\ref{eq2}), this suggests that the rotor angular momentum
$R$ is smaller in the yrast band than in the wobbling band, since
$g_p-g_R=0.78$ is positive. In the following, we will see that this behavior
is induced by a wobbling motion.

In Fig.~\ref{fig1}(a), we also show the $g$-factors as functions of $I$ for the
special cases $\bm{j}_p\perp \bm{I}$, $\bm{j}_p\parallel \bm{I}$, and $\bm{j}_p\perp
\bm{R}$. One can observe that for $\bm{j}_p\parallel \bm{I}$ the $g$-factor is quite
close the results obtained in the yrast and wobbling bands, whereas in the other two
cases it lies far away. This indicates the angular momenta of the proton $\bm{j}_p$
and total nuclear system $\bm{I}$ are oriented almost parallel to each other in the
yrast and wobbling bands. For illustration we present also the $g$-factor according
to Eq.~(\ref{eq2}), with $\langle\bm{R}^2\rangle=R(R+1)$ taking rotor angular
momentum quantum numbers: $R=I-j$, $R=I-j+1$, and $R=I-j+2$. One sees that the curve
with $R=I-j+2$ agrees best with the experimental values at $I=11/2$, which indicates
that the rotor angular momentum quantum number $R$ is close to $R=2$ at the bandhead.
For higher spins $I$ the curves $R=I-j$ and $R=I-j+1$ agree better with experimental
values in the yrast and wobbling bands, respectively.

At $I=11/2$ the calculated SQM $Q(11/2)=-1.29~e\textrm{b}$ comes out close to
upper limit of the experimental value $Q(11/2)=-1.71\pm 0.34~e\textrm{b}$.
We note that the more sophisticated MCSM calculations in
Ref.~\cite{Laskar2020PRC} give $Q(11/2)=-1.25~e\textrm{b}$. This good agreement
indicates that calculations in the PRM have correctly accounted for the
structure of the collective (rotational) states. The calculated SQM decreases
with increasing spin $I$ as a consequence  of the denominator $(I+1)(2I+3)$ in
Eqs.~(\ref{eq10})-(\ref{eq11}).

In Fig.~\ref{fig1}(b) results for the contributions $Q_0(I)$ and $Q_2(I)$ as calculated
by Eqs.~(\ref{eq10}) and (\ref{eq11}) are shown separately. It is found that
$Q_0(I)$ is almost zero, since it gets strongly suppressed by the factor $\cos\gamma
=-0.07$. Hence, $Q_2(I)\simeq Q(I)$ which is shown in the lower part of
Fig.~\ref{fig1}(b) for the following values of total angular components along the
$s$-axis and $l$-axis: $(I_s,I_l)=(I-1/2, 3/2)$, $(I-1, 3/2)$, $(I-3/2, 5/2)$, and
$(I-5/2, 5/2)$. The former two and latter two curves agree with the experimental
results for states in the yrast and wobbling bands, respectively. Clearly,
the $I_s$ values in the yrast band are larger than those in the wobbling band.
We shall see that this feature is again caused by the wobbling motion. Altogether,
the values $Q_2(I) \simeq Q(I)$ are smaller for states in the yrast band than in the
wobbling band (note that $\sin\gamma=-0.99$). At the same time, the negative values
of the SQMs are caused by the fact that the total angular momentum $\bm{I}$ aligns
mainly along the $s$-axis.

In order to better understand the behavior of the $g$-factor and SQM as functions of
the spin $I$, we will discuss in the following the angular momentum geometry of the
proton-particle and the collective rotor.

\begin{figure}[!ht]
  \begin{center}
    \includegraphics[width=8.0 cm]{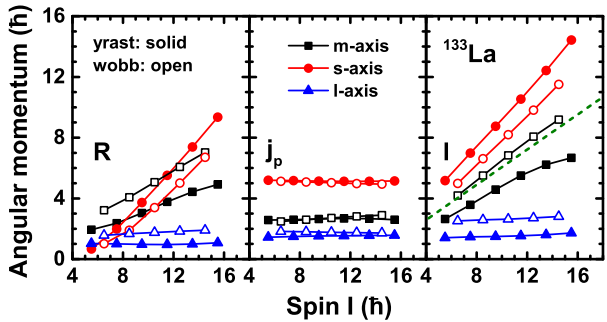}
    \caption{Angular momentum components along the intermediate ($m$-),
    short ($s$-), and long ($l$-) axis of the rotor, proton-particle,
    and total nuclear system as functions of the spin $I$ for the yrast and wobbling
    bands in $^{133}$La. Each component is a root-mean-square expectation value of an
    angular momentum operator, e.g. $I_s=\langle \hat{I}_s^2\rangle^{1/2}$. The dashed
    lines corresponds to the average quantity $\sqrt{I(I+1)/3}$.}\label{fig2}
  \end{center}
\end{figure}

In Fig.~\ref{fig2}, we present the calculated angular momentum components
along the $m$-, $s$-, and $l$-axis of the rotor ($\bm{R}$), proton-particle
($\bm{j}_p$), and total nuclear system ($\bm{I}$) as functions of the spin $I$ for
the yrast and wobbling bands in $^{133}$La. The proton angular momentum $\bm{j}_p$
aligns mainly with the $s$-axis, because its torus-like probability distribution has
a maximal overlap with the triaxial nuclear core in the $ml$-plane~\cite{Frauendorf1996ZPA}.
The $s$-component is constant $j_{s} \simeq 5$ over
the whole spin region for both the yrast and the wobbling band. The $\bm{R}$
favors an alignment in the $sm$-plane with a very small $l$-component,
because $\mathcal{J}_l$ is the smallest. With increasing spin $I$, the
$R_s$ increases faster than $R_m$, due to the gradually increase of
$\mathcal{J}_s$. This behavior of $\bm{R}$ combined with $\bm{j}_p$ gives that
$I_s$ is the largest. The latter feature leads to the negative values of
$Q_2(I)$ as shown in Fig.~\ref{fig1}(b). In addition, the component $I_s$
($I_m$) for states in the yrast band is larger (smaller) than for those in
the wobbling band. A wobbling motion takes place about the $s$-axis, as
it also occurs in the neighboring isotone
$^{135}$Pr~\cite{Matta2015PRL, Sensharma2019PLB}. The additional
alignment of $R_s$, due to the increase of $\mathcal{J}_s$,
stabilizes the wobbling motion about $s$-axis. This fact is consistent
with increasing wobbling energies~\cite{Biswas2019EPJA}. Since $\mathcal{J}_s$
and $\mathcal{J}_m$ are almost equal, a classification of the wobbling mode
as longitudinal or transverse seems inappropriate. The angular momentum
geometry is just more complex, corresponding to an intermediate
situation between the two limits.

Moreover, the component $I_m$ is close to the average quantity $\sqrt{I(I+1)/3}$.
This explains why the contribution $Q_0(I)$ almost vanishes, as shown in
Fig.~\ref{fig1}(b). The component $I_l$ is small, with a value $I_l\approx 3/2$
for the yrast band and $I_l\approx 5/2$ for the wobbling band. This explains
why the outcome of the analytical formula for $Q_2(I)$ in Eq.(14) agrees
so well with the full calculation in the PRM for yrast and wobbling bands.

\begin{figure}[!ht]
  \begin{center}
    \includegraphics[width=8.0 cm]{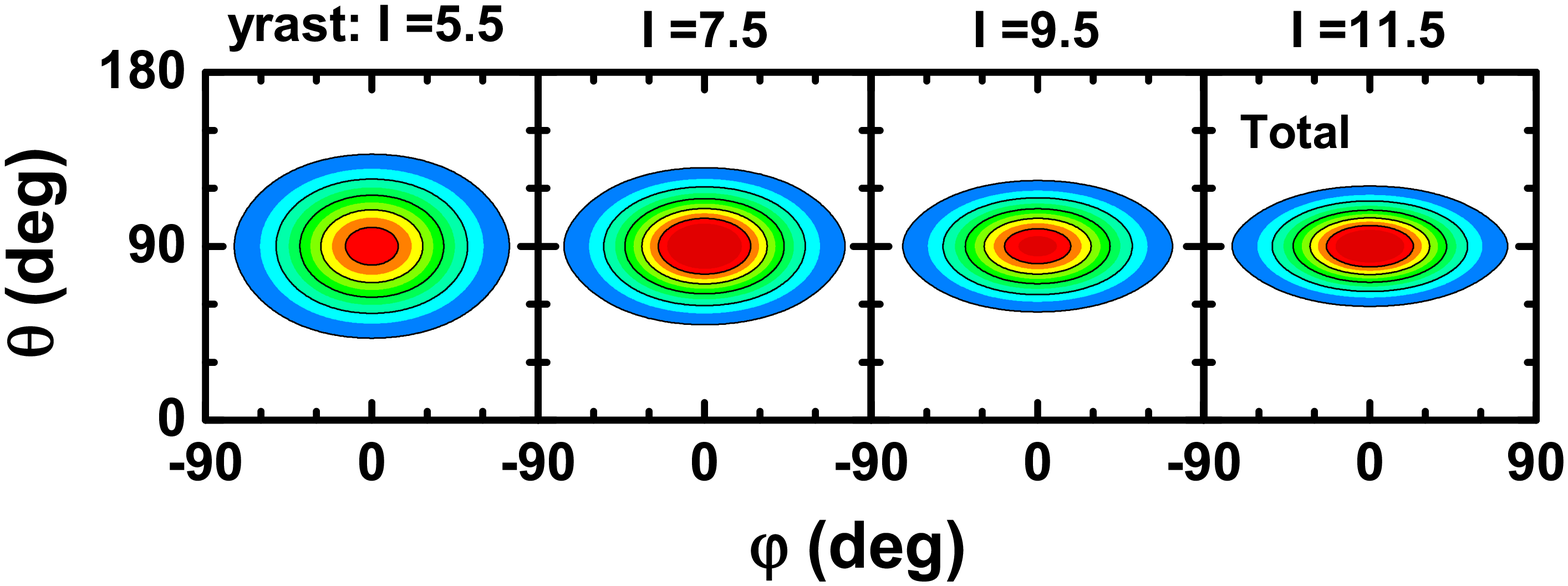}\\
    \includegraphics[width=8.0 cm]{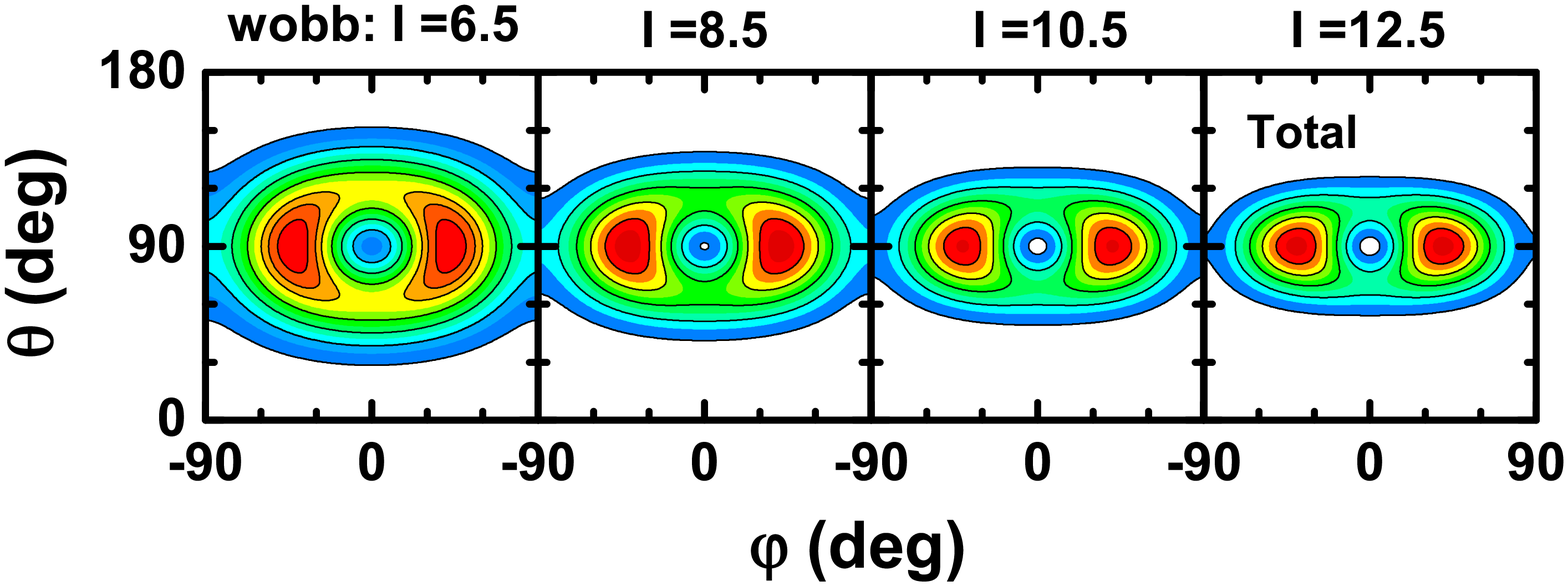}
    \caption{Azimuthal plots, i.e., probability density distributions for the
    orientation of the angular momentum $\bm{I}$ on the $\theta\varphi$-sphere as
    calculated in the  PRM for states in the yrast and wobbling bands of $^{133}$La.}
    \label{fig3}
  \end{center}
\end{figure}

\begin{figure*}[!ht]
  \begin{center}
    \includegraphics[width=7.8 cm]{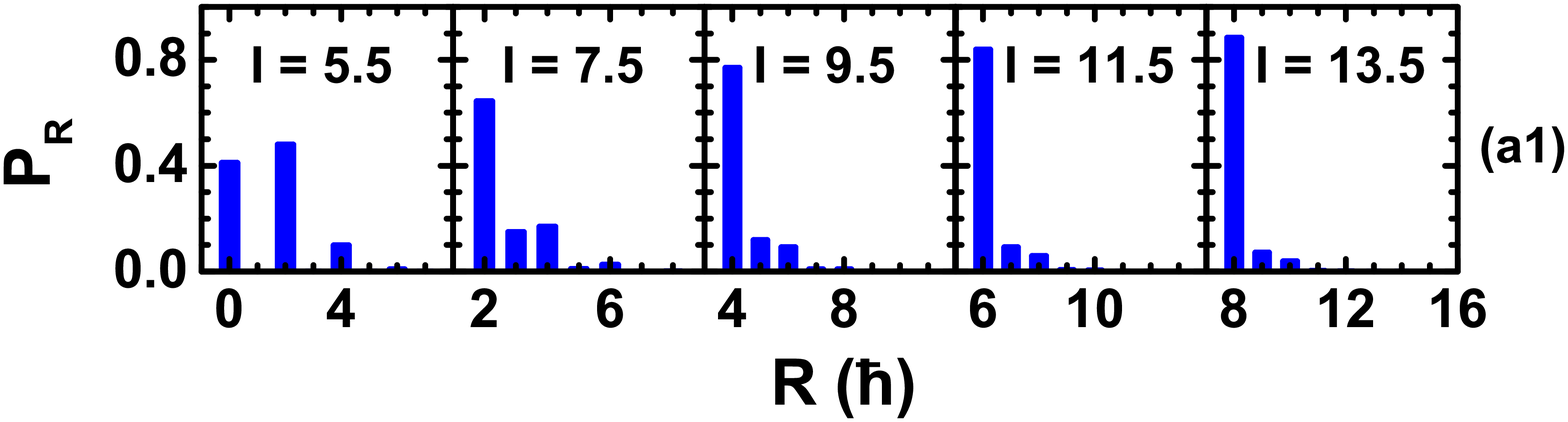}\quad
    \includegraphics[width=7.8 cm]{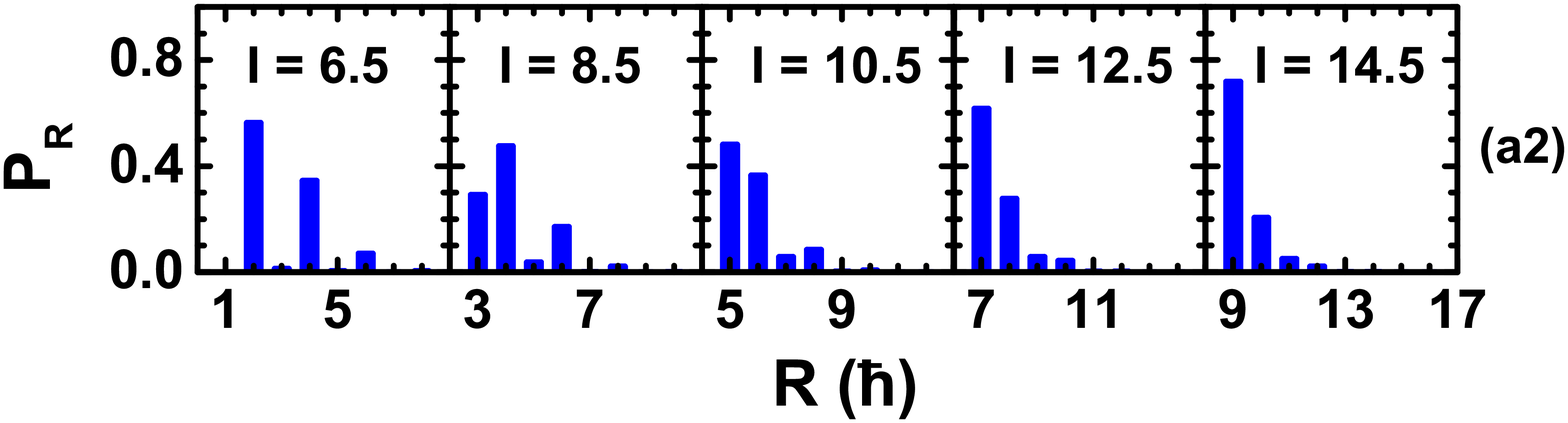}\\
    \includegraphics[width=7.8 cm]{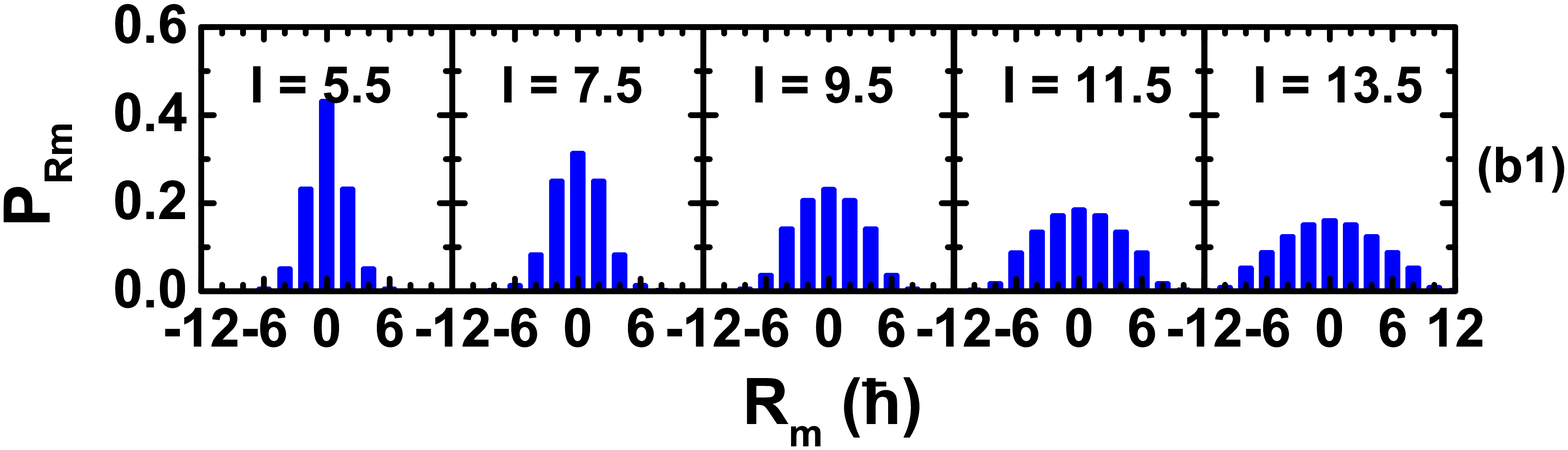}\quad
    \includegraphics[width=7.8 cm]{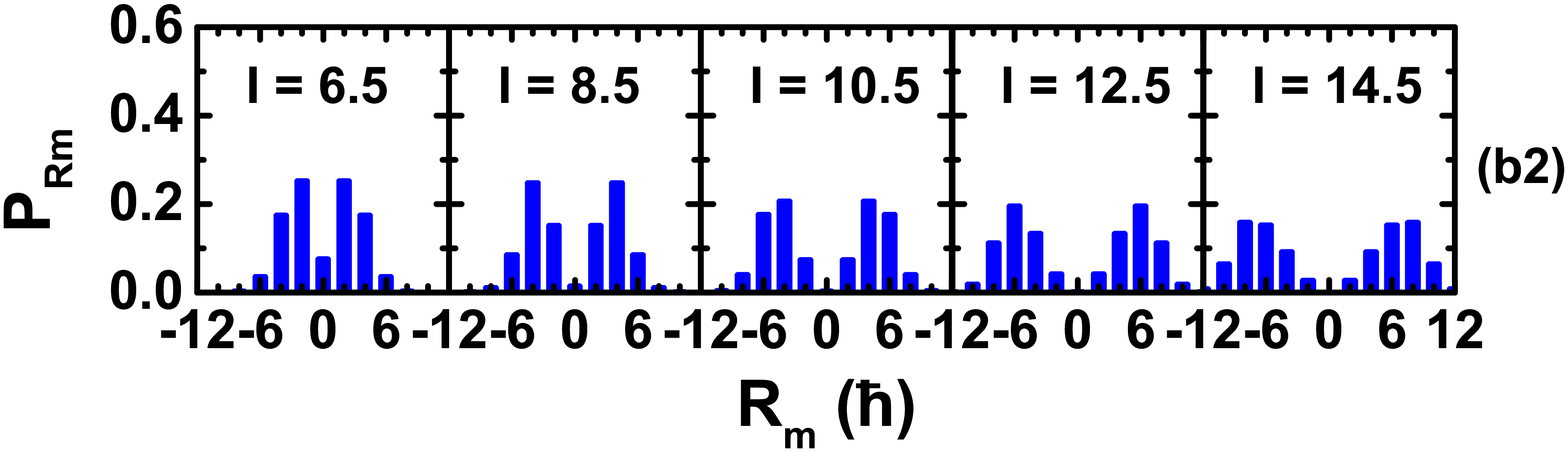}\\
    \includegraphics[width=7.8 cm]{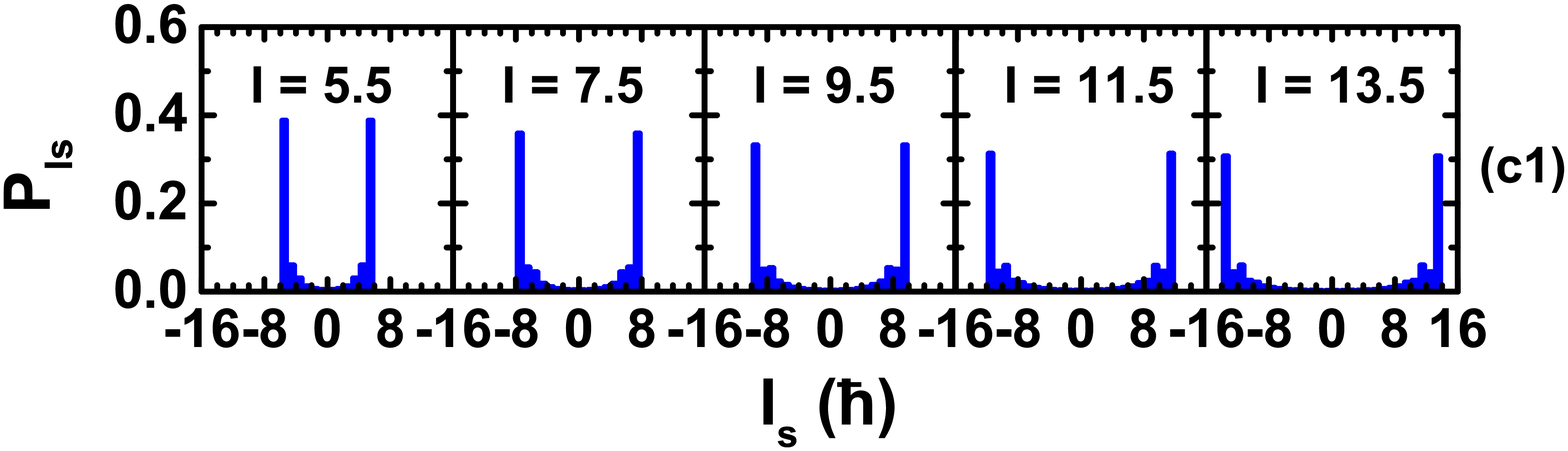}\quad
    \includegraphics[width=7.8 cm]{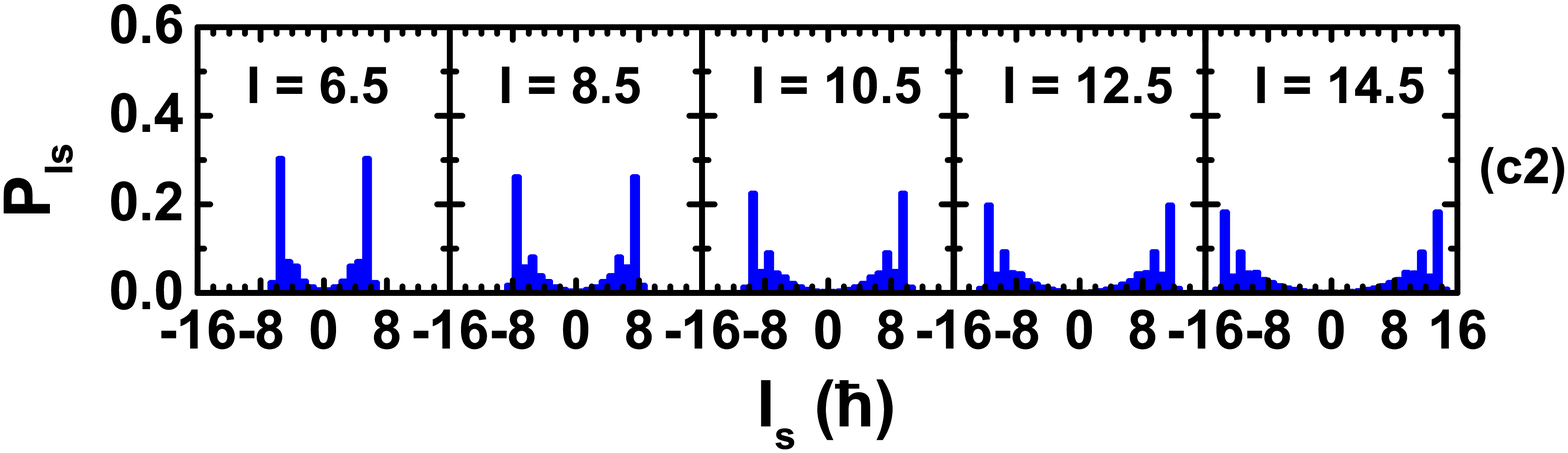}
    \caption{Probability distributions for the rotor angular momentum
    ($R$-plot $P_R$, a1-a2), for the projection of the rotor angular
    momentum onto the $m$-axis ($K_R$-plot $P_{R_m}$, b1-b2), and for the
    projection of the total angular momentum onto the $s$-axis ($K$-plot
    $P_{I_s}$, c1-c2) as calculated in the PRM for states in the yrast and wobbling
    bands of $^{133}$La.}\label{fig4}
  \end{center}
\end{figure*}

In order to illustrate further the wobbling motion about the $s$-axis, the
probability density distributions $\mathcal{P}(\theta,\varphi)$ for the orientation
of the total angular momentum $\bm{I}$ on the $\theta\varphi$-sphere (called azimuthal
plots~\cite{F.Q.Chen2017PRC, F.Q.Chen2018PLB, Q.B.Chen2018PRC_v1, Streck2018PRC})
are shown in Fig.~\ref{fig3} for states in the yrast and wobbling bands of
$^{133}$La. Here, $\theta$ is a polar angle between the total spin $\bm{I}$ and the
$l$-axis, and $\varphi$ is an azimuthal angle in the $sm$-plane measured from the
$s$-axis. Over the entire spin region, the distributions $\mathcal{P}(\theta,\varphi)$
are centered about $\theta=90^\circ$, which corresponds to very small $I_l$-components,
as shown earlier in Fig.~\ref{fig2}. For states in the yrast band the maximum
lies at $\varphi=0^\circ$, which represents a highest probability of aligning
$\bm{I}$ along the $s$-axis. To the contrary, states in the wobbling band
have a minimum at $\varphi=0$. For these wobbling states, the maximal
probability lies on a rim around the minimum, and $\mathcal{P}(\theta,\varphi)$
reflects in this way the wobbling motion (or precession) of $\bm{I}$ about the $s$-axis.
Note that one obtains here precisely the distributions as expected for the wobbling
motion~\cite{Streck2018PRC, Q.B.Chen2019PRC_v1}, namely $\varphi$-symmetric wave
functions for the ($n=0$) yrast band and $\varphi$-antisymmetric wave functions
for the ($n=1$) wobbling band. Moreover, the feature that the distributions
centered at $\varphi=0^\circ$ do not extend out to $\varphi=\pm 90^\circ$
indicates that the wobbling mode in $^{133}$La is very stable, which is
guaranteed by a gradual increase of $\mathcal{J}_s$ with $I$.

In Fig.~\ref{fig4}, we show the calculated probability distributions
for the rotor angular momentum ($R$-plots $P_R$, a1-a2), for the projection
of the rotor angular momentum onto the $m$-axis ($K_R$-plots $P_{R_m}$, b1-b2),
and for the projection of the total angular momentum onto the $s$-axis ($K$-plots
$P_{I_s}$, c1-c2) in the yrast and wobbling bands of $^{133}$La. These detailed
results do further support the picture of a wobbling motion about the $s$-axis.

The $R$-plots (a1) and (a2) show a similar behavior as those for the nucleus
$^{135}$Pr~\cite{Streck2018PRC}, namely, for states in the yrast band $R$ is
almost a good quantum number. The $P_R$-distributions have a pronounced peak at
the minimum value of $R=I-j$, except for the bandhead with $I=11/2$, where the
maximal weight occurs at $R=I-j+2=2$. For states in the wobbling band an
admixture of substates with $R=I-j$ and $R=I-j+1$ is present. An exception occurs
again at the bandhead $I=13/2$, where the peaks lie at $R=I-j+1$ and $I-j+3$.
According to these characteristics, one can understand the behavior of the
$g$-factor as a function of spin $I$, as shown in Fig.~\ref{fig1}(a).

The $K_R$-plots (b1) and (b2) illustrate how the picture of a wobbling oscillation
arises. The distributions $P_{R_m}$ display large admixtures of various
values of $R_m$, which have their origin in the wobbling motion of the rotor towards
to the $m$-axis. At $R_m=0$, the distribution $P_{R_m}$ has a finite value for states
in the yrast band, while it vanishes for states in the wobbling band. This is a
characteristic of the one-phonon excitation of the wobbling mode and it consistent
with the $\varphi$-symmetric wave functions for ($n=0$) yrast states and
$\varphi$-antisymmetric wave functions for ($n=1$) wobbling states, as visualized
by the azimuthal plots $\mathcal{P}(\theta,\varphi)$ in Fig.~\ref{fig3}.

The $K$-plots (c1) and (c2) display that the prominent peaks of the distribution
$P_{I_s}$ appear at $I_s=\pm I$ for states in the yrast band and at $I_s=\pm(I-1)$
for states in the wobbling band. This corresponds to the classical picture of the
wobbling motion. The yrast state with spin $I$ and the neighboring wobbling state
with spin $I+1$ have similar angular momentum components along the $s$-axis. The
total angular momentum in a wobbling state with $I+1$ has to precess (wobble)
with respect to $s$-axis to reach $I_s \simeq I$.

In summary, the $g$-factor and SQM for the wobbling mode of $^{133}$La have
been investigated in the framework of the PRM. The calculation reproduces the
available $g$-factor and SQM data well. The properties of the
$g$-factor and SQM as functions of spin $I$ have been interpreted
by analyzing the angular momentum components of the rotor, proton-particle, and total
nuclear system with the help of various quantum mechanical probability distributions:
Azimuthal plots, $R$-plots, $K_R$-plots, and $K$-plots. It has been
demonstrated that the wobbling mode in $^{133}$La corresponds to a wobbling
of $\bm{I}$ about the $s$-axis. The $g$-factor at the bandhead of yrast band
gives the information $R\simeq 2$ about the rotor angular momentum $\bm{R}$.
The variation of the $g$-factor with spin $I$ indicates that the angular
momenta of the proton-particle and total nuclear system are oriented parallel
to each other. The negative values of SQMs are caused by the fact that the
total angular momentum $\bm{I}$ aligns with the $s$-axis. The differences
of the SQM between states in the yrast and wobbling bands can be traced
back to the wobbling motion. Therefore, the $g$-factor and SQM are good
probes for depicting the picture of wobbling motion. Future experimental
measurements of $g$-factors and the SQMs for states in the high-spin region
are strongly suggested in order to test the theoretical predictions presented
in this work.

\section*{Acknowledgements}

This work has been supported in parts by Deutsche
Forschungsgemeinschaft (DFG) and National Natural Science Foundation
of China (NSFC) through funds provided by the Sino-German CRC 110
``Symmetries and the Emergence of Structure in QCD'' (DFG Grant No.
TRR110 and NSFC Grant No. 11621131001), the US Department of Energy
(Grant No. DE-FG02-95ER40934), the National Key R\&D
Program of China (Contract No. 2017YFE0116700 and No.
2018YFA0404400), the NSFC under Grant No. 11935003, and the State
Key Laboratory of Nuclear Physics and Technology of Peking
University (Grant No. NPT2020ZZ01). The work of U.-G.M. was also
supported by the Chinese Academy of Sciences (CAS) through a
President's International Fellowship Initiative (PIFI) (Grant No.
2018DM0034) and by the VolkswagenStiftung (Grant No. 93562).



\end{document}